\documentclass{aa}

\usepackage[utf8]{inputenc}
\usepackage[varg]{txfonts}
\usepackage{graphicx}
\usepackage{empheq}
\usepackage{natbib}
\bibpunct{(}{)}{;}{a}{}{,} % to follow the A&A style

\begin{document}

\title{Observations of Cygnus X-1 in the MeV band by the INTEGRAL imager}

\author{R. Walter\inst{1}, M. Xu\inst{1,2}}
\institute{$^1$Astronomy Department, University of Geneva, Chemin d’Ecogia 16, 1290 Versoix,  Switzerland\newline
$^2$Institute of High Energy Physics, Chinese Academy of Sciences,19B Yuquan Lu, Shijingshan District, Beijing, 100049, China\newline
$^1$\email{roland.walter@unige.ch}
}

\date{Received July 20, 2016; accepted April 24, 2017}

% \abstract{}{}{}{}{} 
% 5 {} token are mandatory
 
\abstract
  {
The spectrum of the MeV tail detected in the black-hole candidate Cygnus X-1 remains controversial as it appeared much harder when observed with the INTEGRAL Imager IBIS than with the INTEGRAL spectrometer SPI or CGRO. 
  } 
  {
We present an independent analysis of the spectra of Cygnus X-1 observed by IBIS in the hard and soft states. 
  }
  {  
We developed a new analysis software for the PICsIT detector layer and for the Compton mode data of the IBIS instrument and calibrated the idiosyncrasies of the PICsIT front-end electronics.
  }
   {
The spectra of Cygnus X-1 obtained for the hard and soft states with the INTEGRAL imager IBIS are compatible with those obtained with the INTEGRAL spectrometer SPI, with CGRO, and with the models that attribute the MeV hard tail either to hybrid thermal/non-thermal Comptonisation or to synchrotron emission.
   }
   {}

%\authorrunning{}
%\titlerunning{}

\maketitle

\section{Introduction\label{Sect:Introduction}}

Cygnus X-1 is one of the rare accreting black-hole candidates in a high-mass X-ray binary system \citep{1972Natur.235...37W,2011ApJ...742...84O} and certainly the most studied thanks to its exceptionally bright and persistent X-ray emission. The source spends the majority of the time in the hard state where the thermal Compton emission from the corona dominates the weak thermal emission from the accretion disk. In the soft state the disk emission becomes prominent and the thermal coronal emission vanishes \citep{1999ASPC..161..375C,
2004AIPC..714...79C,2002ApJ...578..357Z,2011MNRAS.416.1324Z,2012MNRAS.422.1750Z,2004PThPS.155...99Z}. The radio jet \citep{2001MNRAS.327.1273S} is particularly strong during the hard state \citep{2011MNRAS.416.1324Z}.

Observing the non-thermal gamma-ray emission originating from the corona and from the jet is important to understand the source geometry and probe the accretion-ejection paradigm. In the hard state, this spectral component can only be studied above $\sim 300$ keV. In the soft state the situation is more complex as the thermal Compton emission may peak at higher energies. The observations of CGRO \citep{2002ApJ...572..984M} led to several interpretations on the nature of the MeV emission \citep{1999MNRAS.309..496G,2002ApJ...578..357Z} and several observing campaigns were organised with INTEGRAL to probe it further \citep{2013MNRAS.430..209D}.

While soft gamma-ray polarisation above 300 keV was detected by INTEGRAL \citep{2011Sci...332..438L,2012ApJ...761...27J}, indicating synchrotron emission, the spectral shape of the high-energy component remained controversial as it appeared much harder with the imager IBIS \citep{2011Sci...332..438L,2015ApJ...807...17R} than with the spectrometer SPI or with CGRO \citep{2012ApJ...761...27J,2002ApJ...572..984M}. 

In this paper we revisit the IBIS data obtained in the MeV band, with the PICsIT detector plane and with the Compton events detected by both the ISGRI and PICsIT layers, with independent analysis methods in order to progress solving that controversy.

\cite{2013MNRAS.434.2380M} and \cite{2013ApJ...766...83S} used Fermi and Agile observations of Cyg X-1 in the GeV band and concluded that the average non-thermal gamma-ray component does not extend above 100 MeV but that short flares of yet unknown nature could.

\section{Data analysis\label{Sect:Data}}

\subsection{Picsit analysis\label{Sect:Picsit}}
PICsIT, the PIxellated Cesium Iodide Telescope, is the high-energy lower detector layer of the IBIS coded mask instrument on-board INTEGRAL \citep{2003A&A...411L.149L}. The PICsIT detector is made of 4096 8.5x8.5x30 mm$^3$ CsI crystals, which are sensitive to soft gamma-rays in the range 175 keV and 10 MeV. The angular resolution provided by the coded mask is 12 arcmin. The square field of view is 9x9 square degrees in the fully coded area and extends to 27x27 square degrees in the partially coded area with decreasing sensitivity. The detector is surrounded with lateral and bottom bismuth germanate oxide anti-coincidence shields.

Most PICsIT data are stored in spectral imaging histograms accumulating spectra in 256 energy channels per pixel for periods of typically 20-120 minutes. Different histograms are created for single and multiple events. The latter correspond to photons detected in more than one pixel, in a group of 4x4 pixels sharing the same ASIC chip, or in the same semi-module (256 pixels) by the front end electronics.

The multiple event selection is not uniform over the detector plane \citep{2003A&A...411L.203S,2003A&A...411L.189D} as photons have a higher probability to be detected as multiple events in the centre of the 4x4 pixel groups and as the ASICs are not perfectly synchronised in time. 

\begin{figure}
\centering
\includegraphics[width=0.8\hsize]{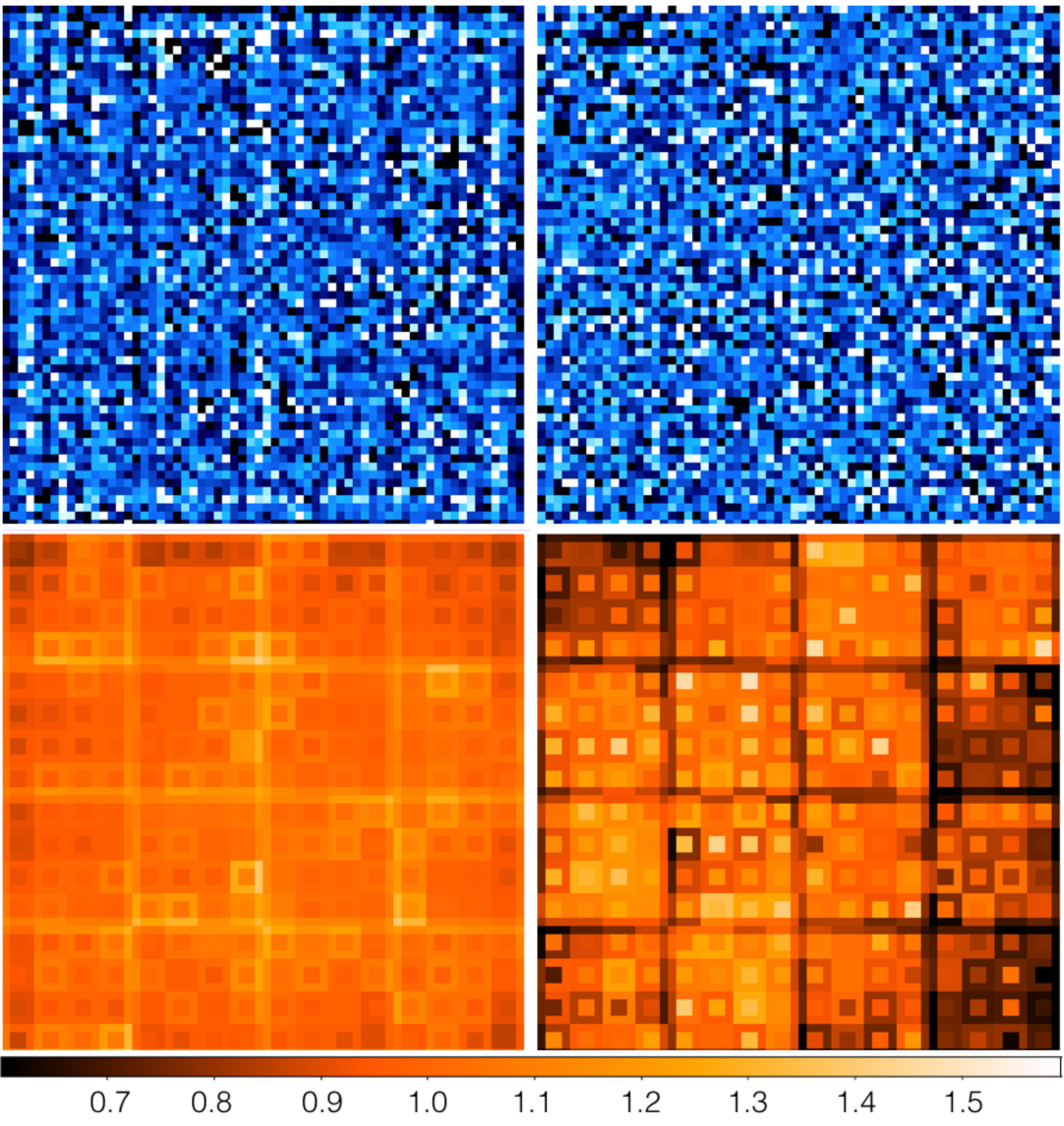}
\caption{Typical background (first row) and uniformity (second row) maps obtained for single (left) and  multiple (right) events in the energy band 336-448 keV. The colour scale is indicated for the uniformity maps. The colour bars of the background maps were chosen to distinguish 90\% of the events. The background maps are rescaled  during the analysis to match the exposure of each pointing.}
\label{fig:picsitub}
\end{figure}

As the PICsIT signal is completely dominated by cosmic-ray induced background and the signal from any astrophysical source is very weak and spread over the detector, uniformity and background maps can be generated by accumulating events over very long time periods. Uniformity maps are multiplied with the detector image to correct for pixel-by-pixel response variation. Background maps are subtracted from the detector image after the appropriate exposure time scaling. We generated uniformity and background maps every 10 spacecraft revolutions (i.e. accumulating about 30 days of data). Multiplicative uniformity parameters were defined for each module and semi-module and between the central and border of the 4x4 pixel groups and fitted to the averaged detector image for each of the eight standard PICsIT energy bands (single: 203-252-336-448-672-1036-1848-3584-6720 keV; multiple: 336-448-672-1036-1848-3584-6720-9072-13440 keV). We found that the uniformity factors vary by a factor of three across the detector plane and are rather constant with time; the time variability is much smaller than the pixel-to-pixel uniformity corrections. The background maps, obtained by dividing the average detector image by the uniformity map, are rather uniform (Fig. \ref{fig:picsitub}). The histogram of the background values are reasonably close to Gaussian distributions except at low energy where a widening of the distribution can be observed, most probably related to accumulation of cosmic-ray tracks \citep{2006NuPhS.150..349L}.

\begin{figure}
\centering
\includegraphics[width=0.8\hsize]{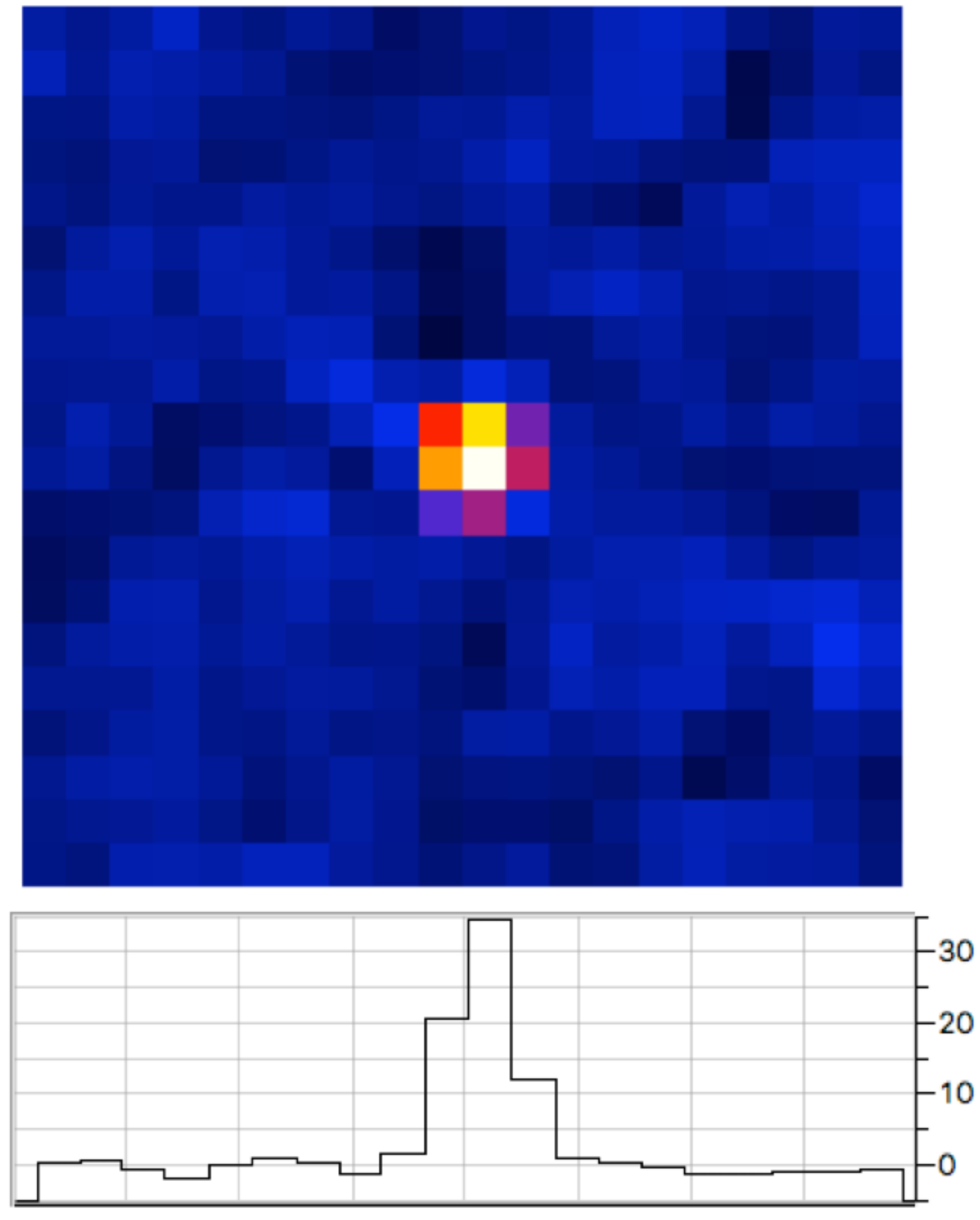}
\caption{Significance image and profile (along the galactic plane axis) of the Crab nebula for revolution 43 (80 ksec exposure) in the second energy band (252-336 keV). This image can be compared to the central region of Figure 3 from \cite{2003A&A...411L.189D}.}
\label{fig:picscrabima}
\end{figure}

Uniformity maps for single and multiple events show opposite trends as the probability of detecting multiple and single events are anti-correlated. Globally, and considering the effect of dithering, the main effect of the detector uniformities is that the probability for a photon to be detected as a single or as a multiple event varies with energy, which has a significant effect on the observed source spectra.

The non-uniformities have an important impact on the quality of the analysis products, in particular for images. We rewrote the PICsIT image deconvolution software including proper handling of the uniformity and background maps based on the algorithms presented by \cite{2003A&A...411L.223G}. Figure \ref{fig:picscrabima} shows the image of the Crab nebula obtained during revolution 43, where the source significance reaches 35 while the standard analysis provides only about 10 \citep{2003A&A...411L.189D}. The RMS of the sky image background is found to be enhanced (RMS $\sim 2\sigma$) in the first energy band only, where the cosmic-ray tracks are predominant.

\subsection{Compton mode analysis\label{Sect:Compton}}
The INTEGRAL Soft Gamma-Ray Imager \citep[ISGRI;][]{2003A&A...411L.141L} detector plane is placed on top of PICsIT. The ISGRI detector is made of 16384 4$\times$4$\times$2 mm$^{3}$ cadmium telluride semiconductor pixels, covering the energy from 15 keV to 1 MeV with an energy resolution of 8\% FWHM around 100 keV. The CdTe layer is made of eight identical modules each hosting $32\times 64$ pixels. The ISGRI detector is ideally suited to build sky images with a good spatial resolution.

Events detected simultaneously by both the ISGRI and PICsIT layers (within a predefined coincidence time window) are tagged on-board as Compton events and transmitted as a photon-by-photon list in the INTEGRAL telemetry. These Compton events are classified as single or multiple according to the number of excited PICsIT pixels. The energy of the event is the sum of that detected by the excited ISGRI and PICsIT pixels and the scattering angle is calculated using the PICsIT pixel with the largest energy deposit.

To decrease the background we implemented several event selection cuts, following the receipt of \cite{2007ApJ...668.1259F}. Only the Compton single events (excluding those from the calibration source) with a total energy between 200 keV and 5 MeV were kept to maximise the accuracy of the Compton scattering reconstruction. The outermost pixels of each of the ISGRI and PICsIT modules (creating specific patterns on the images) were disregarded to remove scattering on the structures separating the detector modules. Of the events, 63\% survived these selections. Events with reconstructed off-axis angle $>15^{\circ}$ (76\% of the above selected events) were disregarded to remove those coming from the back and sides of the detector. 

\begin{figure}
\centering
\includegraphics[width=\hsize]{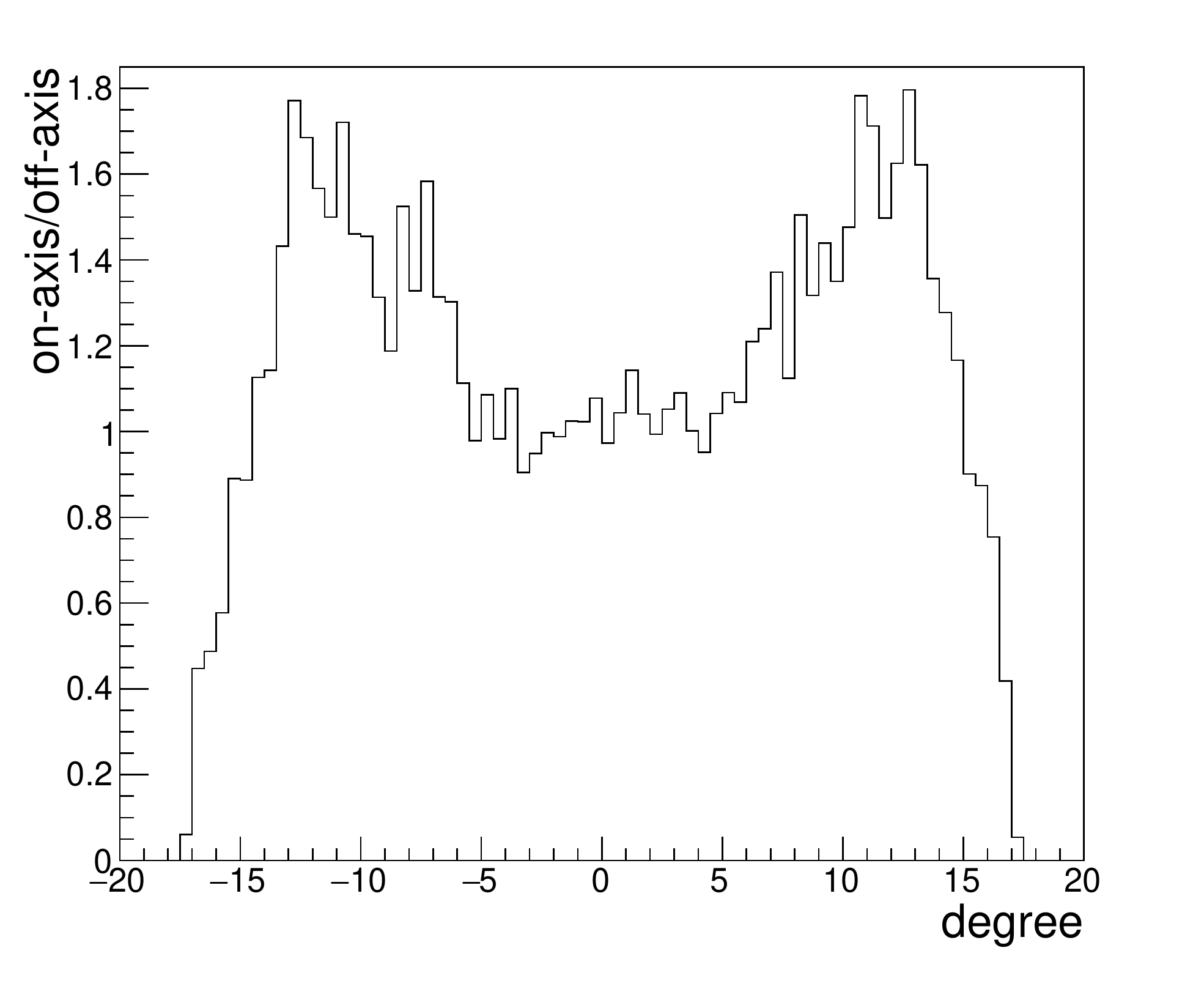}
\caption{Ratio of the Compton event rate obtained for on-axis and off-axis $(10\degr)$ sources, as a function of the offset between the source and the reconstructed direction of each event obtained via the Compton interaction geometry.}
\label{fig:compoffaxis}
\end{figure}

We also tried several additional cuts to keep only events detected from a direction compatible with the source of interest, as implemented by \cite{2007ApJ...668.1259F}, and found that the source signal-to-noise ratio was not significantly improved; the purple/cyan spectra in Fig. \ref{fig:crabspec} were obtained without/with a source direction selection. We therefore decided not to use such source position related event selections, in particular as their results depend on the source off-axis angle. Indeed Fig. \ref{fig:compoffaxis} shows the ratio of the Compton event rates obtained for an on-axis and for a $10\degr$ off-axis source as a function of the offset between the reconstructed direction of an event and the true direction of the source. The results of a cut on this offset therefore depend on the source off-axis angle, making the analysis of the data complex and requiring a set of spectral responses built according to these off-axis angles, which is not available. 

\begin{figure}
\centering
\includegraphics[width=0.8\hsize]{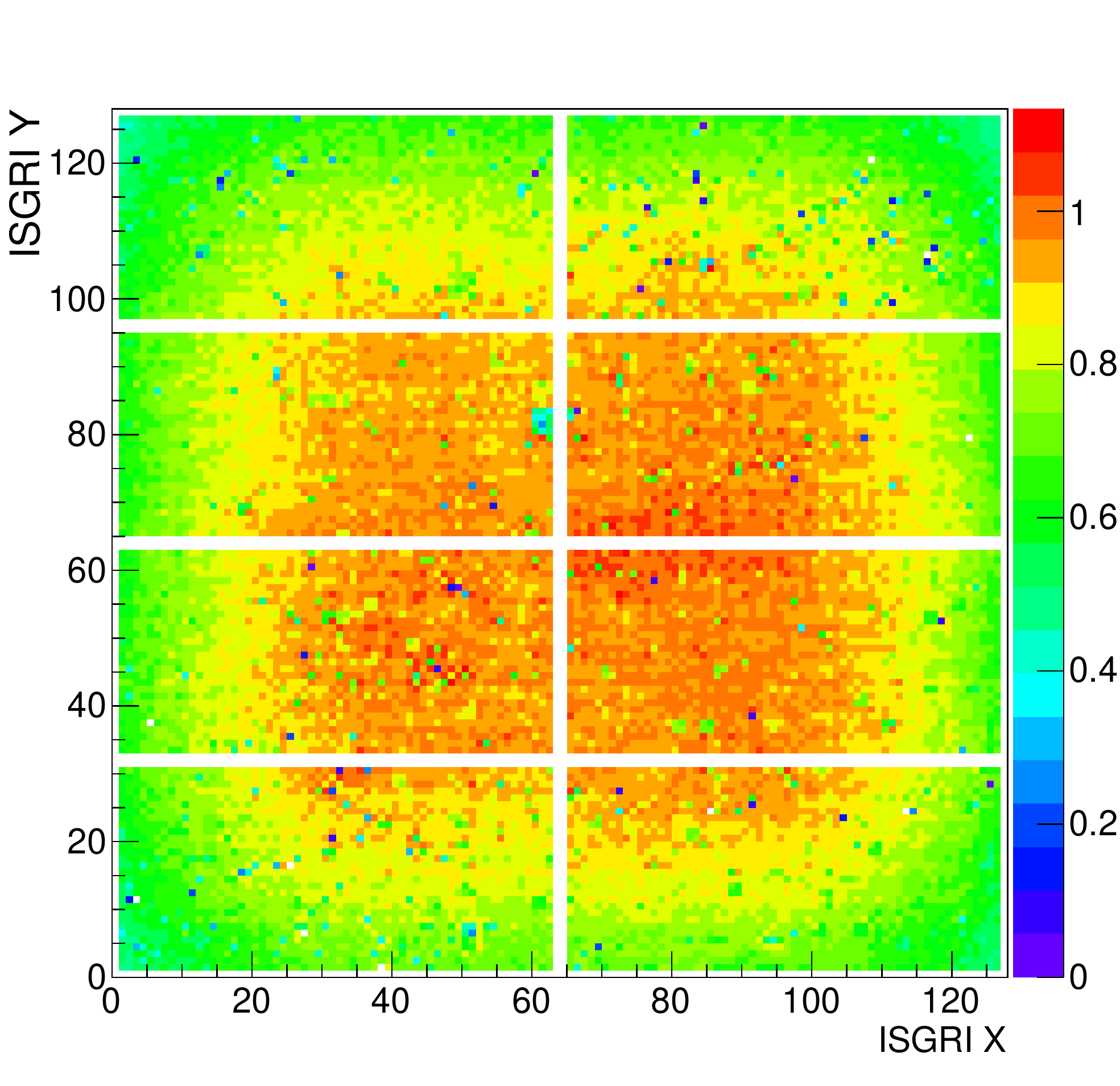}
\caption{Pixel-by-pixel non-uniformity correction map (400-600 keV).}
\label{fig:u-map}
\end{figure}

As events scattering at the edge of the ISGRI layer can miss the PICsIT layer, a higher Compton count rate is observed in the central area of the detector plane. This non-uniformity must be taken into account before  image deconvolution to reduce the coding noise. We built a series of detector non-uniformity images in fixed energy bands defined as the normalised sum of the Compton events detected in each ISGRI pixel over the complete mission (Fig. \ref{fig:u-map}). The detector images are finally divided by these uniformity images.

Because of the high background count rate, random coincidences between uncorrelated ISGRI and PICsIT events within the Compton time window (variable across the mission) are also recorded as Compton events. The contribution of these fake Compton events were evaluated following \cite{2007ApJ...668.1259F}. Detector images were built for both detected and fake Compton events for each pointing and energy band. These detected and fake detector images, corrected for non-uniformities, were then deconvolved to obtain sky images using the algorithms presented in \cite{2003A&A...411L.223G}.

\subsection{Test with the Crab nebula\label{Sect:Crab}}
\begin{figure}
\centering
\includegraphics[width=\hsize]{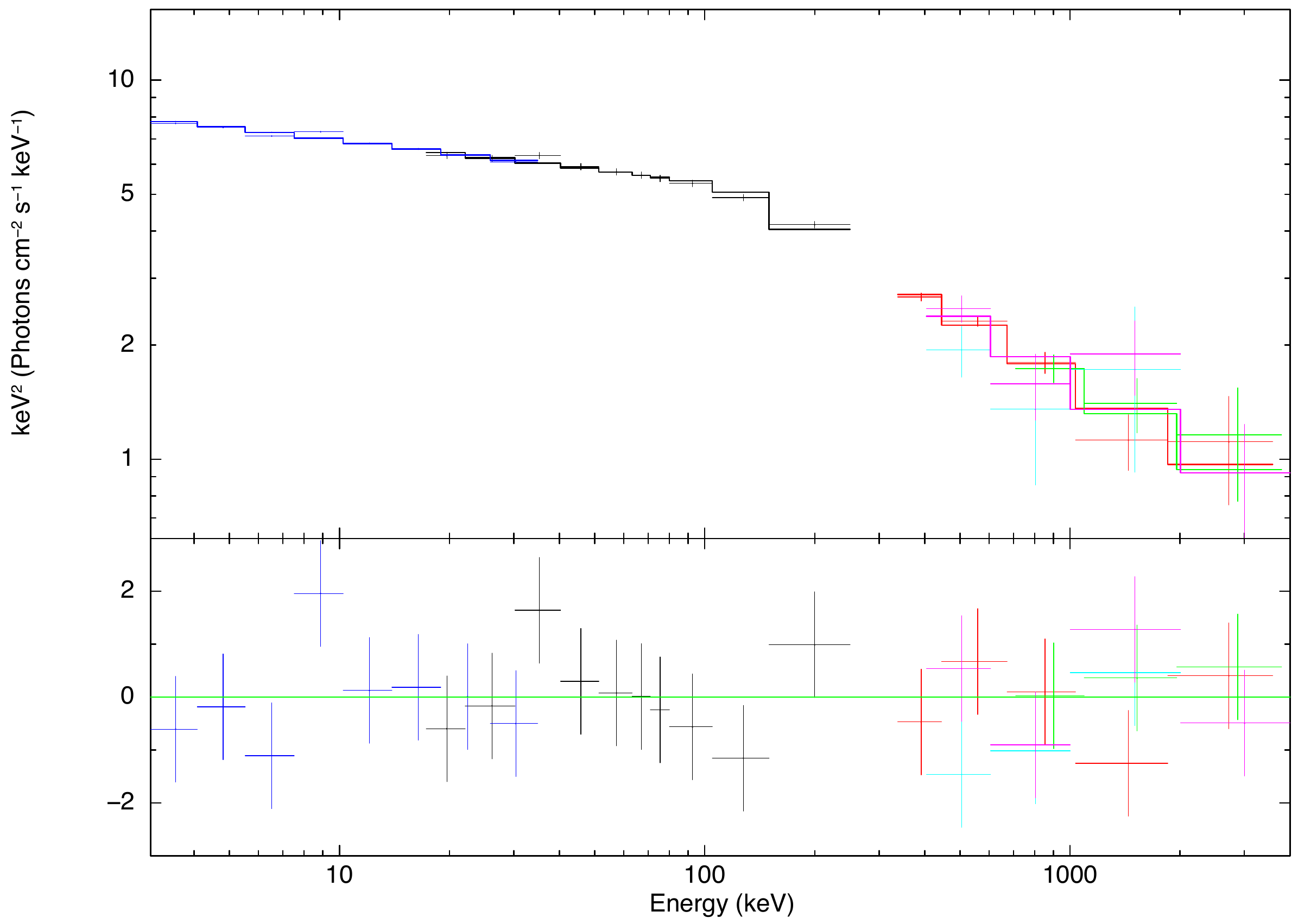}
\caption{Unfolded spectrum of the Crab nebula with residuals as observed by INTEGRAL (instrument/color/normalisation: JEM-X/blue/$0.96\pm 0.04$, ISGRI/black/1.00 (fixed), PICsIT single and multiple/red and green/$1.21\pm 0.15$, Compton mode/purple and cyan/normalisation fixed according to two different event selections) fitted with a broken powerlaw model. Residuals are in units of standard deviations.}
\label{fig:crabspec}
\end{figure}

Source spectra are extracted from the PICsIT mosaic images using  \texttt{mosaic\_spec} and fitted with the response function generated before launch from Monte-Carlo simulations. The PICsIT spectra of the Crab nebula show significant deviations to the expectation mostly because of a lack of single events and an excess of multiple events for energies below 1 MeV, as expected from the observed uniformities. As translating these uniformities in a model of the on-board electronic and generating a new Monte-Carlo model is beyond the scope of our work, we corrected the PICsIT ancillary response file to account for the average effect of the non-uniformities, reducing the response for single events and increasing it for multiple events for all energy bins. 

Mosaics of the detected and fake Compton sky images were generated and the detected and fake source images were fitted as above to obtain count rates in all energy bands. The source spectrum was obtained subtracting the spectrum resulting from the fake events from that obtained from the detected events. Fig. \ref{fig:crabspec} shows two Compton mode spectra that were obtained with (purple) and without (cyan) an event selection based on the source position. Their cross-calibration constants were fixed according to the expectations for the respective event cuts. The signal-to-noise ratio of these two spectra are similar.

The resulting spectrum of the Crab nebula as observed by JEM-X, ISGRI, PICsIT, and the Compton mode on-board INTEGRAL is shown in Fig. \ref{fig:crabspec} fitted with a single broken power-law model with best fit parameters $\Gamma_1=2.11\pm0.02$, $\Gamma_2=2.55\pm0.15$, E$_{\rm break} =116\pm30$ keV, and N$_{\rm 1 keV}=8.9\pm0.9$ ph/(keV cm$^2$ s). The effective exposures are $\approx 1.5\times 10^6$ s for JEM-X and $\approx 3\times 10^6$ s for IBIS (and the Compton mode). The cross-calibration constants, indicated in the caption of Fig. \ref{fig:crabspec}, are reasonably consistent in view of the variability of the Crab nebula and calibration uncertainties. The goodness of fit $\chi^2_\nu$ is 0.77 assuming systematic uncertainties of 2\% (as recommended to fit ISGRI data) and 1.00 using systematics of 1.5\%.

The high-energy spectral slope is significantly steeper than the value obtained by INTEGRAL SPI \citep{2009ApJ...704...17J}. As this difference might be an artefact of the PICsIT response generated before launch, we built an alternate PICsIT ancillary response file to match the spectral parameters obtained with the spectrometer and will use it to compare our results with those of \cite{2015ApJ...807...17R}. With this alternate response, the flux at 1 MeV is about 50\% larger than shown in Fig. \ref{fig:crabspec}.

\subsection{Data selection and analysis\label{Sect:Selection}}
In order to obtain results on Cyg X-1 that could be compared with other analyses, we selected the data as \cite{2015ApJ...807...17R}, who provided us with the list of INTEGRAL pointings corresponding to the hard and soft states of Cygnus X-1 as defined in their analysis. This selection was based on hardness ratios and intensity measurements obtained by RXTE/ASM and Swift/BAT observations made within less than six hours of the INTEGRAL pointings \citep{2013A&A...554A..88G}.

We performed the JEM-X, ISGRI, PICsIT, and Compton mode analyses for the selected data to derive the broad-band spectra of Cygnus X-1 in the soft and hard states. Because of their different fields of view, the effective exposures obtained with JEM-X are about two times shorter than those obtained with IBIS. The spectral normalisations obtained for the two instruments are therefore different as the observations are not truly simultaneous.

\section{Results and discussion\label{Sect:Results}}
\begin{figure}
\centerline{
\includegraphics[width=\hsize]{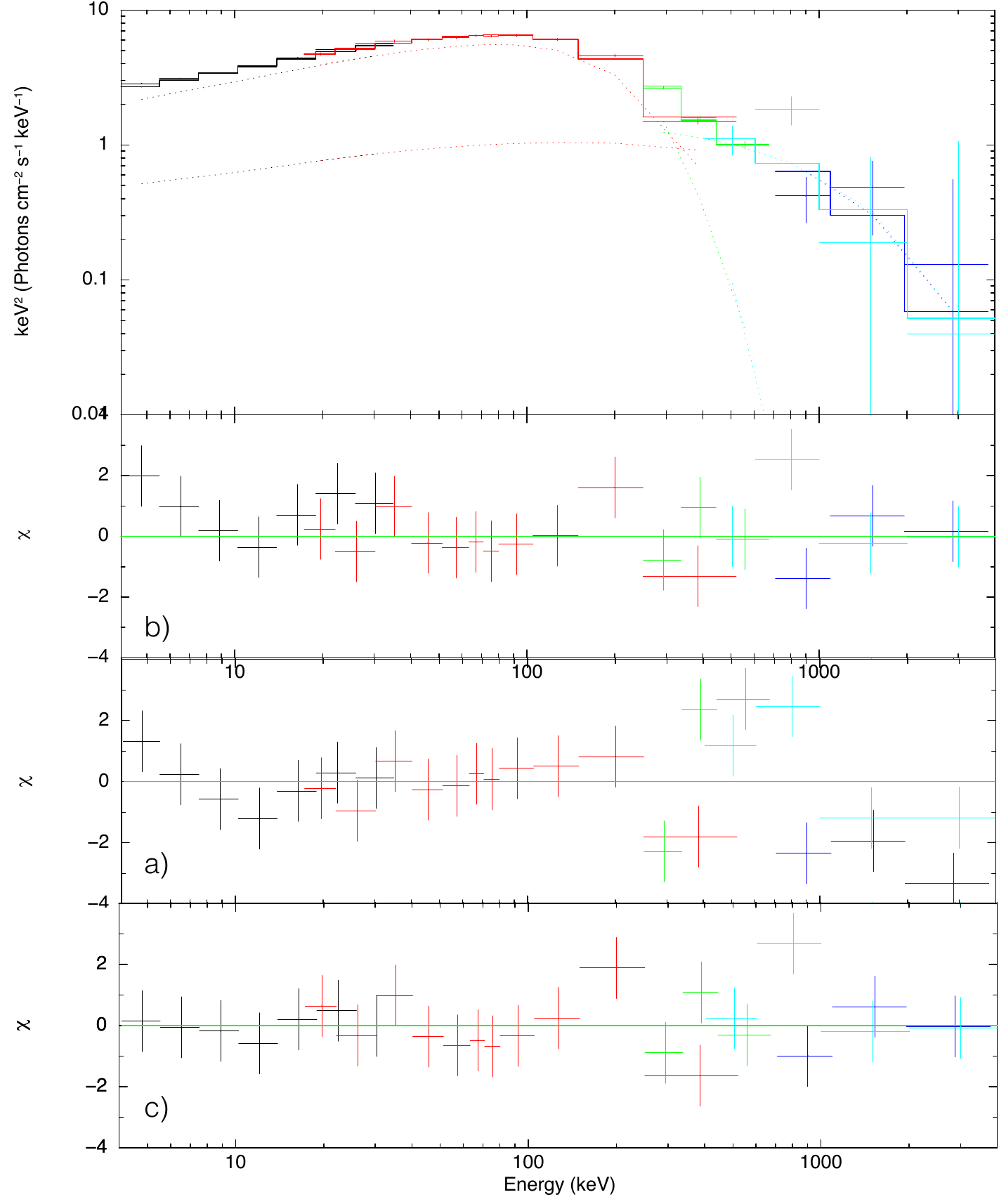}
}

\caption{Top panel: unfolded spectrum of the hard state of Cygnus X-1 using model (b), i.e. reflection (comptt + cutoff power-law). Following panels: residuals to the fits in unit of standard deviations for models (b), (a, reflection (comptt+ power-law)), and (c, eqpair). The corresponding model parameters are listed in Tab. \ref{tab:param}. The data are from JEM-X (black), ISGRI (red), PICsIT (green and blue), and the Compton mode (cyan).}
\label{fig:spectra}
\end{figure}

\begin{figure}
\centerline{
\includegraphics[width=\hsize]{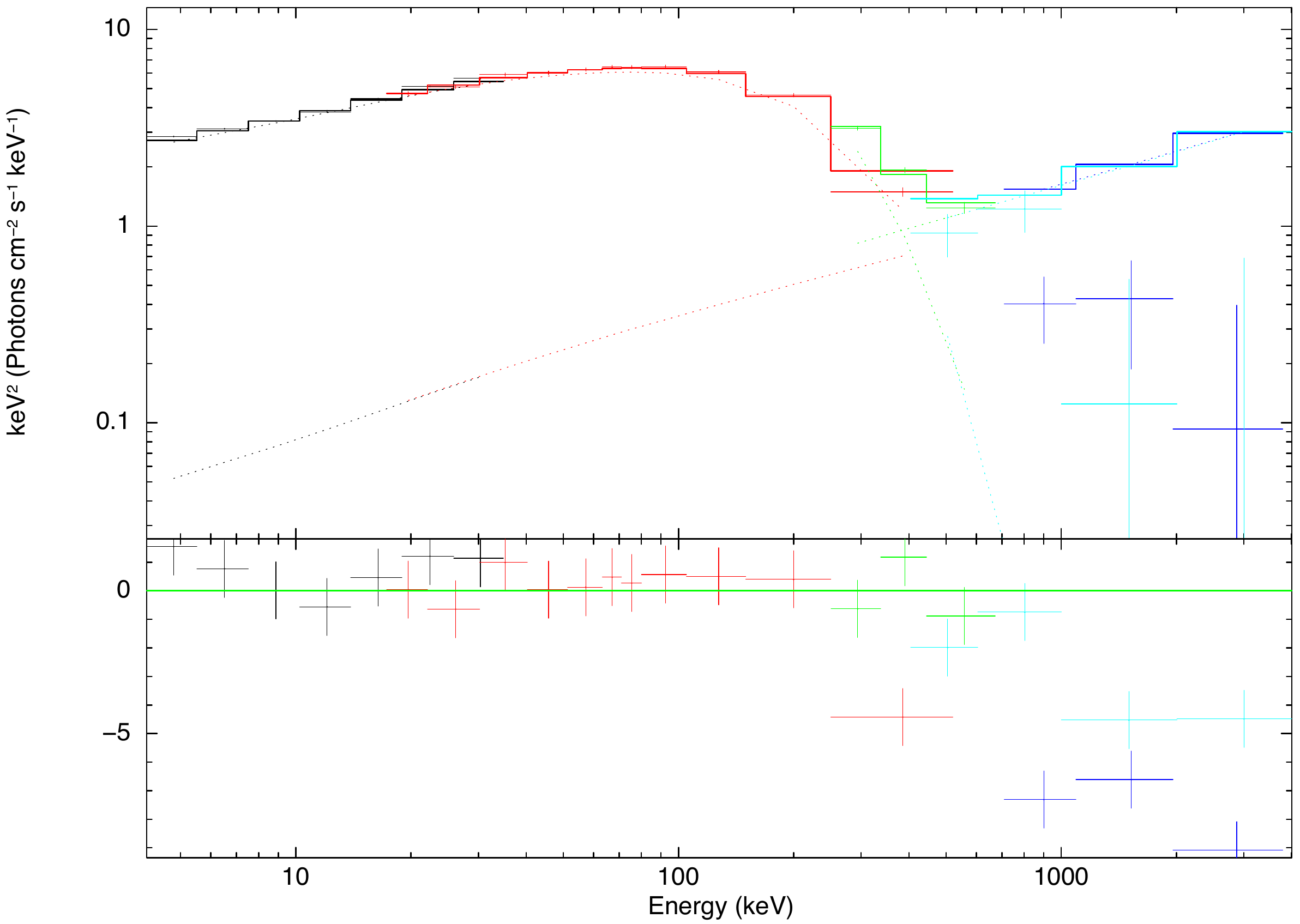}
}

\caption{Unfolded spectrum and residuals for the hard state of Cygnus X-1 using model (d) presented by \cite{2015ApJ...807...17R}. The data are the same as in Fig. \ref{fig:spectra}. The alternate PICsIT response, introduced at the end of Sect. \ref{Sect:Crab}, was used here for spectral unfolding.}
\label{fig:rodriguez}
\end{figure}

\begin{table}
\caption{Spectral models and parameters. The parameter $\eta$ is the covering factor of the reflection and N are the normalisation of the various spectral components. The thermal comptonisation is characterised by the corona electron temperature kT$_{\rm e}$ and optical depth $\tau$. The eqpair model is characterised by the injected electron spectral slope $\Gamma$ and by the soft, hard, non-thermal, and thermal compactness, the other parameters being fixed to values usually used for Cyg X-1. The parameter C is the PICsIT/ISGRI cross-calibration constant resulting from the fit, the cross-calibration factor of the Compton mode are fixed according to the events selection. The hard power-law flux is given in the range 0.4-1 MeV. The goodness of fit of the first model (limited by the absence of a high-energy cutoff) does not allow us to obtain reliable uncertainties. The results shown as model d) refer to the fit of the hard state spectrum obtained by \cite{2015ApJ...807...17R}, which could be compared to model a).}
\label{tab:param}
\begin{center}
\begin{tabular}{p{3.5cm}p{4cm}}
\hline\noalign{\smallskip}
\hline\noalign{\smallskip}
\noalign{\smallskip}
Spectrum \& Model & Parameters\\
\noalign{\smallskip}
\hline\noalign{\smallskip}
Hard state\newline a) refl(comptt+pow)  & 
$\chi^2_\nu=2.7$\newline 
C = 1.27\newline
$\eta = 0.15$\newline
kT$_{\rm e} = 48.6$ keV\newline
$\tau = 1.15$\newline
N$_{comp} = 0.08$\newline
$\Gamma = 1.6$\newline
N$_{pow} = 0.04$ ph/(keV cm$^2$ s)
F$_{pow} = 6.3\ 10^{-10}$ erg/(cm$^2$ s)
\\
\hline\noalign{\smallskip}
Hard state\newline b) refl(comptt+cutoffp) &
$\chi^2_\nu=1.0$\newline 
C = 1.27\newline
$\eta = 0.18\pm0.2$\newline
kT$_{\rm e} = 40.0\pm2.1 $ keV\newline
$\tau = 1.49\pm0.08$\newline
N$_{comp} = 0.07$\newline
$\Gamma = 1.86\pm0.2$\newline
E$_{\rm cut} = 0.7\pm0.3$ MeV\newline
N$_{cutp} = 0.7$  ph/(keV cm$^2$ s)
F$_{pow} = 9.7\ 10^{-10}$ erg/(cm$^2$ s)
\\
\hline\noalign{\smallskip}
Hard state\newline c) eqpair &
$\chi^2_\nu=1.0$\newline 
C = 1.26\newline
$\eta = 0.2\pm0.1$\newline
$\Gamma_{\rm inj} = 2.3\pm0.6$\newline
l$_{\rm bb} = 21\pm 6$\newline
l$_{\rm h}$/l$_{\rm s} = 6.7\pm 0.6$\newline
l$_{\rm nt}$/l$_{\rm h} = 0.14\pm 0.08$\newline
N$ = (4.0\pm0.2)\times 10^{-8}$
\\
\hline\noalign{\smallskip}
\multicolumn{2}{l}{Hard state fit from \cite{2015ApJ...807...17R}}\\
d) refl\ (comptt+pow) &
$\eta = 0.13\pm0.02$\newline
kT$_{\rm e} = 53\pm2$ keV\newline
$\tau = 1.15\pm0.04$\newline
$\Gamma = 1.4^{+0.2}_{-0.3}$\newline
F$_{pow} = 19\ 10^{-10}$ erg/(cm$^2$ s)
\\
\hline\noalign{\smallskip}
Soft state\newline e) eqpair &
$\chi^2_\nu=1.4$\newline 
C = 1.12\newline
$\eta = 1$ (fixed)\newline
$\Gamma_{\rm inj} = 3.4\pm0.2$\newline
l$_{\rm bb} = 1.2\pm1.0$\newline
l$_{\rm h}$/l$_{\rm s} = 0.31\pm 0.09$\newline
l$_{\rm nt}$/l$_{\rm h} = 0.8\pm 0.3$\newline
N$ = (7\pm1)\times 10^{-7}$
\\
\hline\noalign{\smallskip}
\\
\end{tabular}
\end{center}
\end{table}

Figure \ref{fig:spectra} shows the hard state spectrum of Cygnus X-1 fitted with different models, using the original PICsIT response. The parameters are listed in table \ref{tab:param}.

The hard state spectrum was first modelled with a thermal comptonisation plus a hard power-law tail (model a), similar to that used by \cite{2011Sci...332..438L} and \cite{2015ApJ...807...17R}. The best fit parameters of the hard tail are dominated by the low energy $(<300$ keV) data and do not provide a good representation (at a level of $5\sigma$) of the upper limits above 400 keV. 

A cutoff power-law model for the hard tail (model b, and the top panel of Fig. \ref{fig:spectra}) provides a much better representation of our data. Our results, which were obtained independently for the PICsIT and Compton mode data, and our best fit parameters are consistent with those obtained by \cite{2012ApJ...761...27J} using data from the INTEGRAL spectrometer SPI, with those presented by \cite{2012MNRAS.423..663Z}, based on a completely independent analysis of the PICsIT data, and finally with the CGRO results \citep{2002ApJ...572..984M}. The PICsIT/ISGRI cross-calibration constants (see Tab. \ref{tab:param}) obtained from the spectral adjustments are very similar to those obtained on the Crab nebula (Sect. \ref{Sect:Crab}). Our hard state spectrum is also in agreement with the model of \cite{2014MNRAS.440.2238Z} attributing the MeV hard tail either to a hybrid Comptonisation tail or to synchrotron emission, the latter being favoured by the detection of polarisation by \cite{2012ApJ...761...27J}.

We have also used the model obtained by \cite{2015ApJ...807...17R} (model d, Fig. \ref{fig:rodriguez}) using the alternate PICsIT response, built to match the expected Crab spectral slope (see end of Sect. \ref{Sect:Crab}), to minimise the discrepancies between that model and our data. This model is clearly too hard and too bright at high energy when compared to the results of our analysis (by more than $20\sigma$). Using the original PICsIT response leads to even larger discrepancies.

In the soft state we could detect Cyg X-1 up to about 500 keV in agreement with the results obtained by the INTEGRAL spectrometer SPI at a slightly lower flux level \citep{2014ApJ...789...26J}. We have not detected an inflection point at 250 keV as suggested by \cite{2014ApJ...789...26J} but this might be related to the thermal Comptonisation model used by these authors to represent the  data at low energies. Our spectrum is also fairly comparable to the CGRO spectrum of the soft state presented by \cite{2002ApJ...572..984M}. Our upper limits above 1 MeV are significantly below those obtained with Comptel. 

The eqpair\footnote{http://www.astro.yale.edu/coppi/eqpair/} model provides an acceptable fit to the spectra of the hard and soft states (models c \& e) with parameters in reasonable agreement with those obtained by \cite{2002ApJ...572..984M}. 

Although we are using the same data, the results presented by \cite{2015ApJ...807...17R} for the hard state are not consistent with ours nor with those of \cite{2012ApJ...761...27J}, \cite{2012MNRAS.423..663Z}, or \cite{2002ApJ...572..984M}. The uncertainties obtained by \cite{2015ApJ...807...17R} are much smaller than ours, the spectral shape is different and its normalisation brighter by $20 \sigma$. These differences are probably related to the subtraction of fake Compton events. We cannot exclude that different Compton event selection may play a role, although we have used less strict event selection cuts.

Further observations of Cygnus X-1 in the hard state by INTEGRAL with PICsIT and the Compton mode can help better constrain the spectral shape of the source beyond 1 MeV. A much more sensitive instrument, such as e-Astrogam \citep{2016SPIE.9905E..2NT}, is required to probe the jet contribution in the soft state and to measure the relation between the inner jet and the disk through variability.

%\begin{acknowledgements}
%\end{acknowledgements}

\bibliographystyle{aa} % style aa.bst
\bibliography{cygx1} % your references Yourfile.bib

\begin{thebibliography}{30}
\expandafter\ifx\csname natexlab\endcsname\relax\def\natexlab#1{#1}\fi

\bibitem[{{Coppi}(2004)}]{2004AIPC..714...79C}
{Coppi}, P. 2004, in American Institute of Physics Conference Series, Vol. 714,
  X-ray Timing 2003: Rossi and Beyond, ed. P.~{Kaaret}, F.~K. {Lamb}, \& J.~H.
  {Swank}, 79--88

\bibitem[{{Coppi}(1999)}]{1999ASPC..161..375C}
{Coppi}, P.~S. 1999, in Astronomical Society of the Pacific Conference Series,
  Vol. 161, High Energy Processes in Accreting Black Holes, ed. J.~{Poutanen}
  \& R.~{Svensson}, 375

\bibitem[{{Del Santo} {et~al.}(2013){Del Santo}, {Malzac}, {Belmont},
  {Bouchet}, \& {De Cesare}}]{2013MNRAS.430..209D}
{Del Santo}, M., {Malzac}, J., {Belmont}, R., {Bouchet}, L., \& {De Cesare}, G.
  2013, \mnras, 430, 209

\bibitem[{{Di Cocco} {et~al.}(2003){Di Cocco}, {Caroli}, {Celesti}, {Foschini},
  {Gianotti}, {Labanti}, {Malaguti}, {Mauri}, {Rossi}, {Schiavone},
  {Spizzichino}, {Stephen}, {Traci}, \& {Trifoglio}}]{2003A&A...411L.189D}
{Di Cocco}, G., {Caroli}, E., {Celesti}, E., {et~al.} 2003, \aap, 411, L189

\bibitem[{{Forot} {et~al.}(2007){Forot}, {Laurent}, {Lebrun}, \&
  {Limousin}}]{2007ApJ...668.1259F}
{Forot}, M., {Laurent}, P., {Lebrun}, F., \& {Limousin}, O. 2007, \apj, 668,
  1259

\bibitem[{{Gierli{\'n}ski} {et~al.}(1999){Gierli{\'n}ski}, {Zdziarski},
  {Poutanen}, {Coppi}, {Ebisawa}, \& {Johnson}}]{1999MNRAS.309..496G}
{Gierli{\'n}ski}, M., {Zdziarski}, A.~A., {Poutanen}, J., {et~al.} 1999,
  \mnras, 309, 496

\bibitem[{{Goldwurm} {et~al.}(2003){Goldwurm}, {David}, {Foschini}, {Gros},
  {Laurent}, {Sauvageon}, {Bird}, {Lerusse}, \&
  {Produit}}]{2003A&A...411L.223G}
{Goldwurm}, A., {David}, P., {Foschini}, L., {et~al.} 2003, \aap, 411, L223

\bibitem[{{Grinberg} {et~al.}(2013){Grinberg}, {Hell}, {Pottschmidt},
  {B{\"o}ck}, {Nowak}, {Rodriguez}, {Bodaghee}, {Cadolle Bel}, {Case}, {Hanke},
  {K{\"u}hnel}, {Markoff}, {Pooley}, {Rothschild}, {Tomsick}, {Wilson-Hodge},
  \& {Wilms}}]{2013A&A...554A..88G}
{Grinberg}, V., {Hell}, N., {Pottschmidt}, K., {et~al.} 2013, \aap, 554, A88

\bibitem[{{Jourdain} \& {Roques}(2009)}]{2009ApJ...704...17J}
{Jourdain}, E. \& {Roques}, J.~P. 2009, \apj, 704, 17

\bibitem[{{Jourdain} {et~al.}(2014){Jourdain}, {Roques}, \&
  {Chauvin}}]{2014ApJ...789...26J}
{Jourdain}, E., {Roques}, J.~P., \& {Chauvin}, M. 2014, \apj, 789, 26

\bibitem[{{Jourdain} {et~al.}(2012){Jourdain}, {Roques}, {Chauvin}, \&
  {Clark}}]{2012ApJ...761...27J}
{Jourdain}, E., {Roques}, J.~P., {Chauvin}, M., \& {Clark}, D.~J. 2012, \apj,
  761, 27

\bibitem[{{Labanti} {et~al.}(2003){Labanti}, {Di Cocco}, {Ferro}, {Gianotti},
  {Mauri}, {Rossi}, {Stephen}, {Traci}, \& {Trifoglio}}]{2003A&A...411L.149L}
{Labanti}, C., {Di Cocco}, G., {Ferro}, G., {et~al.} 2003, \aap, 411, L149

\bibitem[{{Labanti} {et~al.}(2006){Labanti}, {Marisaldi}, \&
  {Segreto}}]{2006NuPhS.150..349L}
{Labanti}, C., {Marisaldi}, M., \& {Segreto}, A. 2006, Nuclear Physics B
  Proceedings Supplements, 150, 349

\bibitem[{{Laurent} {et~al.}(2011){Laurent}, {Rodriguez}, {Wilms}, {Cadolle
  Bel}, {Pottschmidt}, \& {Grinberg}}]{2011Sci...332..438L}
{Laurent}, P., {Rodriguez}, J., {Wilms}, J., {et~al.} 2011, Science, 332, 438

\bibitem[{{Lebrun} {et~al.}(2003){Lebrun}, {Leray}, {Lavocat}, {Cr{\'e}tolle},
  {Arqu{\`e}s}, {Blondel}, {Bonnin}, {Bou{\`e}re}, {Cara}, {Chaleil}, {Daly},
  {Desages}, {Dzitko}, {Horeau}, {Laurent}, {Limousin}, {Mathy}, {Mauguen},
  {Meignier}, {Molini{\'e}}, {Poindron}, {Rouger}, {Sauvageon}, \&
  {Tourrette}}]{2003A&A...411L.141L}
{Lebrun}, F., {Leray}, J.~P., {Lavocat}, P., {et~al.} 2003, \aap, 411, L141

\bibitem[{{Malyshev} {et~al.}(2013){Malyshev}, {Zdziarski}, \&
  {Chernyakova}}]{2013MNRAS.434.2380M}
{Malyshev}, D., {Zdziarski}, A.~A., \& {Chernyakova}, M. 2013, \mnras, 434,
  2380

\bibitem[{{McConnell} {et~al.}(2002){McConnell}, {Zdziarski}, {Bennett},
  {Bloemen}, {Collmar}, {Hermsen}, {Kuiper}, {Paciesas}, {Phlips}, {Poutanen},
  {Ryan}, {Sch{\"o}nfelder}, {Steinle}, \& {Strong}}]{2002ApJ...572..984M}
{McConnell}, M.~L., {Zdziarski}, A.~A., {Bennett}, K., {et~al.} 2002, \apj,
  572, 984

\bibitem[{{Orosz} {et~al.}(2011){Orosz}, {McClintock}, {Aufdenberg},
  {Remillard}, {Reid}, {Narayan}, \& {Gou}}]{2011ApJ...742...84O}
{Orosz}, J.~A., {McClintock}, J.~E., {Aufdenberg}, J.~P., {et~al.} 2011, \apj,
  742, 84

\bibitem[{{Rodriguez} {et~al.}(2015){Rodriguez}, {Grinberg}, {Laurent},
  {Cadolle Bel}, {Pottschmidt}, {Pooley}, {Bodaghee}, {Wilms}, \&
  {Gouiff{\`e}s}}]{2015ApJ...807...17R}
{Rodriguez}, J., {Grinberg}, V., {Laurent}, P., {et~al.} 2015, \apj, 807, 17

\bibitem[{{Sabatini} {et~al.}(2013){Sabatini}, {Tavani}, {Coppi}, {Pooley},
  {Del Santo}, {Campana}, {Chen}, {Evangelista}, {Piano}, {Bulgarelli},
  {Cattaneo}, {Colafrancesco}, {Del Monte}, {Giuliani}, {Giusti}, {Longo},
  {Morselli}, {Pellizzoni}, {Pilia}, {Striani}, {Trifoglio}, \&
  {Vercellone}}]{2013ApJ...766...83S}
{Sabatini}, S., {Tavani}, M., {Coppi}, P., {et~al.} 2013, \apj, 766, 83

\bibitem[{{Stephen} {et~al.}(2003){Stephen}, {Caroli}, {Malizia}, {Natalucci},
  {Bassani}, \& {Di Cocco}}]{2003A&A...411L.203S}
{Stephen}, J.~B., {Caroli}, E., {Malizia}, A., {et~al.} 2003, \aap, 411, L203

\bibitem[{{Stirling} {et~al.}(2001){Stirling}, {Spencer}, {de la Force},
  {Garrett}, {Fender}, \& {Ogley}}]{2001MNRAS.327.1273S}
{Stirling}, A.~M., {Spencer}, R.~E., {de la Force}, C.~J., {et~al.} 2001,
  \mnras, 327, 1273

\bibitem[{{Tatischeff} {et~al.}(2016){Tatischeff}, {Tavani}, {von Ballmoos},
  {Hanlon}, {Oberlack}, {Aboudan}, {Argan}, {Bernard}, {Brogna}, {Bulgarelli},
  {Bykov}, {Campana}, {Caraveo}, {Cardillo}, {Coppi}, {De Angelis}, {Diehl},
  {Donnarumma}, {Fioretti}, {Giuliani}, {Grenier}, {Grove}, {Hamadache},
  {Hartmann}, {Hernanz}, {Isern}, {Kanbach}, {Kiener}, {Kn{\"o}dlseder},
  {Labanti}, {Laurent}, {Limousin}, {Longo}, {Marisaldi}, {McBreen}, {McEnery},
  {Mereghetti}, {Mirabel}, {Morselli}, {Nakazawa}, {Peyr{\'e}}, {Piano},
  {Pittori}, {Sabatini}, {Stawarz}, {Thompson}, {Ulyanov}, {Walter}, {Wu},
  {Zdziarski}, \& {Zoglauer}}]{2016SPIE.9905E..2NT}
{Tatischeff}, V., {Tavani}, M., {von Ballmoos}, P., {et~al.} 2016, in
  \procspie, Vol. 9905, Society of Photo-Optical Instrumentation Engineers
  (SPIE) Conference Series, 99052N

\bibitem[{{Webster} \& {Murdin}(1972)}]{1972Natur.235...37W}
{Webster}, B.~L. \& {Murdin}, P. 1972, \nat, 235, 37

\bibitem[{{Zdziarski}(2012)}]{2012MNRAS.422.1750Z}
{Zdziarski}, A.~A. 2012, \mnras, 422, 1750

\bibitem[{{Zdziarski} \& {Gierli{\'n}ski}(2004)}]{2004PThPS.155...99Z}
{Zdziarski}, A.~A. \& {Gierli{\'n}ski}, M. 2004, Progress of Theoretical
  Physics Supplement, 155, 99

\bibitem[{{Zdziarski} {et~al.}(2012){Zdziarski}, {Lubi{\'n}ski}, \&
  {Sikora}}]{2012MNRAS.423..663Z}
{Zdziarski}, A.~A., {Lubi{\'n}ski}, P., \& {Sikora}, M. 2012, \mnras, 423, 663

\bibitem[{{Zdziarski} {et~al.}(2002){Zdziarski}, {Poutanen}, {Paciesas}, \&
  {Wen}}]{2002ApJ...578..357Z}
{Zdziarski}, A.~A., {Poutanen}, J., {Paciesas}, W.~S., \& {Wen}, L. 2002, \apj,
  578, 357

\bibitem[{{Zdziarski} {et~al.}(2011){Zdziarski}, {Skinner}, {Pooley}, \&
  {Lubi{\'n}ski}}]{2011MNRAS.416.1324Z}
{Zdziarski}, A.~A., {Skinner}, G.~K., {Pooley}, G.~G., \& {Lubi{\'n}ski}, P.
  2011, \mnras, 416, 1324

\bibitem[{{Zdziarski} {et~al.}(2014){Zdziarski}, {Stawarz}, {Pjanka}, \&
  {Sikora}}]{2014MNRAS.440.2238Z}
{Zdziarski}, A.~A., {Stawarz}, {\L}., {Pjanka}, P., \& {Sikora}, M. 2014,
  \mnras, 440, 2238

\end{thebibliography}

\end{document}